\documentclass[a4paper,twosides]{article}
\usepackage{multicol,multirow,graphicx,graphics,fancyhdr,epstopdf}
\usepackage{amsmath,amsfonts,amssymb,bm,upgreek,mathrsfs,mathcomp}
\usepackage[pagewise,switch,columnwise]{lineno}
\usepackage[compress,nospace]{cite}
\usepackage[dvipsnames]{xcolor}
\usepackage{CPL-2023}
\newcommand{\cplyear}{2026} \newcommand{\cplvol}{XX}
\newcommand{\cplno}{x} \newcommand{\cplpagenumber}{xxxxxx}
 \newcommand{\cplpage}{\cplpagenumber-\thepage}

\begin{document}

\vspace* {-4mm} \begin{center}
%-------------------------Title----------------------------
\large\bf{\boldmath{Where Does Tracing of Cosmic Ray in Real Atmosphere Terminate?}}
%------------------------Footnote--------------------------
\footnotetext{\hspace*{-5.4mm}$^{*}$Corresponding authors. Email: huor@iat.cn

\noindent\copyright\,{\cplyear}
\href{http://www.cps-net.org.cn}{Chinese Physical Society} and
\href{http://www.iop.org}{IOP Publishing Ltd}}
\\[5mm]
%------------------------Authors----------------------------
\normalsize \rm{}Du-Xin Zheng$^{1}$, Long Chen$^{1,2}$, Ran Huo$^{1*}$
%----------------------COM. or University-------------------
\\[3mm]\small\sl $^{1}$Particle Physics Research Center,\\ Shandong Institute of Advanced Technology, Jinan 250103, China

$^{2}$Institute for Advanced Technology, \\ Shandong University, Jinan 250061, China
%------------------------Received date----------------------
\\[4mm]\normalsize\rm{}(Received xxx; accepted manuscript online xxx)
\end{center}

%----------------------Abstract and PACS--------------------
\vskip 1.5mm

\small{\narrower In backtracing simulations, which are widely employed to determine cosmic-ray particle trajectories in the geomagnetic field, the atmosphere is typically approximated as an artificial sharp boundary at some low altitude where the traced trajectory terminates. In this paper, we extend beyond this simplified assumption and investigate two realistic physical processes that terminate cosmic-ray particle propagation in the atmosphere: Bethe-Bloch energy loss mechanisms and hard scattering interactions with atmospheric atoms using total cross sections based on the Glauber-Gribov formalism. The former mechanism dominates at low rigidities (for protons below $\sim0.57$~GV), while the latter becomes dominant at higher rigidities. Consequently, we introduce two dimensionless variables up to detailed numerical criteria: the relative rigidity shift due to Bethe-Bloch effects ($\Delta\mathfrak{R}/\mathfrak{R}$), and the expected number of hard scattering events ($\langle N\rangle$). Using the corrected US Standard Atmosphere 1976 model, we demonstrate that the altitude dependence can be factorized as approximately $\exp(-0.14h/\textrm{km})$. Additionally, the effect of the local curvature radius of the trajectory near perigee can be similarly factorized. Our calculations indicate that the simplified sharp-boundary altitude should be at least $50$~km with $\Delta\mathfrak{R}/\mathfrak{R}+\langle N\rangle\lesssim1$ for protons, increasing by more than $15$~km for heavy nuclei such as iron.

\par}\vskip 3mm
\normalsize\noindent{\narrower{DOI: \href{http://dx.doi.org/10.1088/0256-307X/\cplvol/\cplno/\cplpagenumber}{10.1088/0256-307X/\cplvol/\cplno/\cplpagenumber}}

\par}\vskip 5mm

Most human space activity is still confined to low-Earth orbit (LEO). Cosmic rays that penetrate the magnetosphere and the overlying atmosphere create a complex radiation environment~\cite{1,2}. For high-energy instruments operating in LEO, such as AMS-02, theoretical models must decide whether an observed event belongs to the primary cosmic-ray population, i.e.\ whether its rigidity exceeds the directional geomagnetic cutoff~\cite{3,4} (the same technique is used for radiation-dose estimates). This classification is performed numerically by backtracing: a Runge-Kutta integrator propagates the charged-particle trajectory backward from the measured arrival position and direction under the Lorentz force alone~\cite{5}. If the trajectory ultimately escapes the magnetosphere and recedes to infinity, the particle is flagged as primary (usually of galactic origin). Because the magnetic force is always perpendicular to the velocity, it does no work and the particle energy remains constant along the backward path.

Possible termination of a backtraced trajectory occurs near its perigee. Once the particle enters the dense lower atmosphere it interacts with air nuclei and the backward propagation must stop. The key question is how to exactly model such termination. Present practice imposes a sharp boundary in altitude: 20~km (radiation-exposure studies~\cite{6}), 40~km (space-based cosmic-ray experiments~\cite{3}), or 100~km (the K\'{a}rm\'{a}n line, recently adopted by AMS-02 for a conservative galactic-cosmic-ray selection~\cite{2,7,8}). These altitudes are, however, ad hoc. The actual cessation of a particle's trajectory should be dictated by physical processes, namely the microscopic interactions between the cosmic-ray particle and atmospheric nuclei.

Two microscopic mechanisms can terminate the backward extension of a trajectory. The first is a single hard scattering that deflects the incident cosmic-ray particle out of the path elastically or inelastically. The second is Bethe-Bloch energy loss: the continuous slowdown violates the energy-conservation premise on which backtracing relies. Hard scattering is routinely studied at high energies through the air-shower signature recorded by ground-based indirect detectors such as the Large High Altitude Air Shower Observatory (LHAASO)~\cite{9}, the Telescope Array (TA)~\cite{10} and the Pierre Auger Observatory~\cite{11}. However, to the best of our knowledge, no quantitative treatment suited to backward tracking from LEO has yet been presented, in which at the GeV range relevant for backtracing the necessary cross-section data must be extracted from accelerator measurements.

To illustrate the concept and estimate the order-of-magnitude effect, we perform a controlled parametric study in which the two physical processes are evaluated along designed trajectories that neglect geomagnetic bending. A simple, factorized scaling result is derived that can readily replace the ad-hoc sharp boundary whenever a binary stop/no-stop criterion is sufficient. In parallel, the same processes can be integrated along the true curved trajectories produced by numerical backtracing and compared to the same criterion; we reserve this fully consistent treatment for a future publication.

\medskip

{\it Models.} Bethe-Bloch energy loss~\cite{12,13,14} arises from the cumulative effect of a continuous series of binary Coulomb scatterings. The stopping power is
\begin{equation}
\frac{dE}{dx}= -\frac{e^{4}}{8\pi m_{e}u}\frac{Z_{\text{P}}^{2}}{\beta^{2}}\sum_{\text{T}}\frac{\rho_{\text{T}}(h)\,Z_{\text{T}}}{A_{\text{T}}}
\bigg(\ln\Big(\frac{I_{\text{P,max}}^{2}}{I_{\text{T,min}}^{2}}\Big)-\beta^{2}\bigg),
\label{dEOdx}
\end{equation}
where $e$ is the elementary charge, $m_{e}$ the electron mass, and $u$ the unified atomic mass unit. The projectile velocity in units of $c$ is $\beta=v/c$ (we adopt natural units $\hbar=1$ and $c=1$), and $Z_{\text{P}}$ is its atomic number. For each target element T, $\rho_{\text{T}}(h)$ is the mass density contributed by that element at altitude $h$, and $Z_{\text{T}}$, $A_{\text{T}}$ are its atomic and mass numbers. $I_{\text{T,min}}$ is the mean excitation energy that sets the maximum impact parameter for individual collisions; we use the values given in Table 1.1 of Ref.\cite{13}. The maximum transferable energy, corresponding to the minimum impact parameter, is
\begin{equation}
I_{\text{P,max}}=\frac{2m_{e}v^{2}}{1-\beta^{2}}\bigg/\sqrt{1+\frac{2m_{e}}{m_{\text{P}}\sqrt{1-\beta^{2}}}+2\big(\frac{m_{e}}{m_{\text{P}}}\big)^{2}},
\label{Imax}
\end{equation}
with $m_{\text{P}}$ the projectile mass.

In a magnetic field the local gyroradius is $r=\mathfrak{R}/B$, where the rigidity $\mathfrak{R} = p/q = p/(Z_\text{P}e)$ encodes both particle species and kinematics and therefore completely determines the trajectory in backtracing. Using $E^2=p^2+m^2$ ($EdE=pdp$) and $\beta=p/E$, the integrated rigidity loss is
\begin{equation}
\Delta\mathfrak{R}=\int dx\frac{d\mathfrak{R}}{dx}=\int \frac{dx}{\beta Z_\text{P}e}\frac{dE}{dx}.
\label{DR1}
\end{equation}
Throughout the numerical work we treat $\beta$ as constant rather than decreasing along the path\footnote{In the limit $\Delta\mathfrak{R}\to 0$ we also have $\Delta\beta\to 0$, so the approximation is self-consistent. For appreciable $\Delta\mathfrak{R}$ the particle is slowed to velocities near the Bragg peak and the trajectory ends; only in a narrow band around this region is a finite-$\Delta\beta$ treatment required. Incorporating the full velocity decrease along each curved path will be carried out in future work that goes beyond the present rescaling approximation.}, i.e.\ the slowing-down is included only perturbatively. Then the fractional rigidity loss reads
\begin{equation}
\frac{\Delta\mathfrak{R}}{\mathfrak{R}}\approx\frac{e^4}{8\pi m_\textrm{e}u}\frac{Z^2_\textrm{P}}{E\beta^4}\int dx~\sum_\textrm{T}\frac{\rho_\textrm{T}(h)Z_\textrm{T}}{A_\textrm{T}}\bigg(\ln\Big(\frac{I_\textrm{P,max}^2}{I_\textrm{T,min}^2}\Big)-\beta^2\bigg).
\label{DR2}
\end{equation}
This crude estimate clearly fails when $\Delta\mathfrak{R}/\mathfrak{R}\gtrsim 1$. However, in the following we will see that the eventual interested $\Delta\mathfrak{R}/\mathfrak{R}\ll1$ can be obtained by multiplying such results with additional small factors from perigee altitude and curvature radius, in such cases this perturbative calculation holds conveniently. Note that the integrand depends on the projectile only through $m_{\text{P}}$ inside the logarithm and via $\beta=\mathfrak{R}Ze/((\mathfrak{R}Ze)^2+m_{\text{P}}^2)^{1/2}$. If we evaluate the integral once for protons and neglect the weak $m_{\text{P}}$ dependence, the result for any ion scales as $Z_\text{P}^2/(E\beta^4)$. In this approximation a single backtracing performed for protons suffices; the fractional rigidity shift for other species is obtained by scaling.

\medskip

For single hard scattering we use the Glauber-Gribov formalism to obtain proton- and nucleus-nucleus cross sections from the underlying nucleon-nucleon data, adopting the simplifications of Ref.\cite{15}. The total (subscript ``tot'' omitted) hadron-nucleus and nucleus-nucleus cross sections are~\cite{15}
\begin{eqnarray}
\sigma_\textrm{hA}&=&2\pi R_\textrm{hA}^2(A)\ln\Big(1+\frac{A\sigma_\textrm{hN}}{2\pi R_\textrm{hA}^2(A)}\Big),\\
\sigma_\textrm{AA}&=&2\pi\big(R_\textrm{AA}^2(A_\textrm{P})+R_\textrm{AA}^2(A_\textrm{T})\big)\ln\Big(1+\frac{A_\textrm{P}A_\textrm{T}\sigma_\textrm{NN}}{2\pi (R_\textrm{AA}^2(A_\textrm{P})+R_\textrm{AA}^2(A_\textrm{T}))}\Big),
\label{sigmaA}
\end{eqnarray}
with the nuclear radii
\begin{eqnarray}
R_\textrm{hA}(A)&=&\left\{\begin{array}{ll}
r_0A^{1/3}(1+0.1\exp(\frac{A-20}{20})), &A\leq20,\smallskip\\
r_0A^{1/3}(0.8+0.2\exp(\frac{20-A}{20})), &A>20,
\end{array}\right.\\
R_\textrm{AA}(A)&=&\left\{\begin{array}{ll}
r_0A^{1/3}(1-A^{-2/3}), &A<50,\smallskip\\
r_0A^{0.27}, &A\geq50,
\end{array}\right.
\label{RA}
\end{eqnarray}
where $r_{0}=1.1$~fm. The nucleon-nucleon level cross section is weighted from the proton-proton, neutron-neutron and proton-neutron\footnote{The difference between the neutron-proton cross section and the proton-neutron cross section is proportional to the relatively mass difference between proton and neutron, and will be ignored.} cross section
\begin{equation}
A_\textrm{P}A_\textrm{T}\sigma_\textrm{NN}=Z_\textrm{P}Z_\textrm{T}\sigma_\textrm{pp}+(A_\textrm{P}-Z_\textrm{P})(A_\textrm{T}-Z_\textrm{T})\sigma_\textrm{nn}
+\big(Z_\textrm{P}(A_\textrm{T}-Z_\textrm{T})+(A_\textrm{P}-Z_\textrm{P})Z_\textrm{T}\big)\sigma_\textrm{pn},
\label{sigmaN}
\end{equation}
and for proton-nucleon scattering we set $Z_\textrm{P}=A_\textrm{P}=1$.

\vskip 4mm

\fl{1}\centerline{\includegraphics[width=6cm]{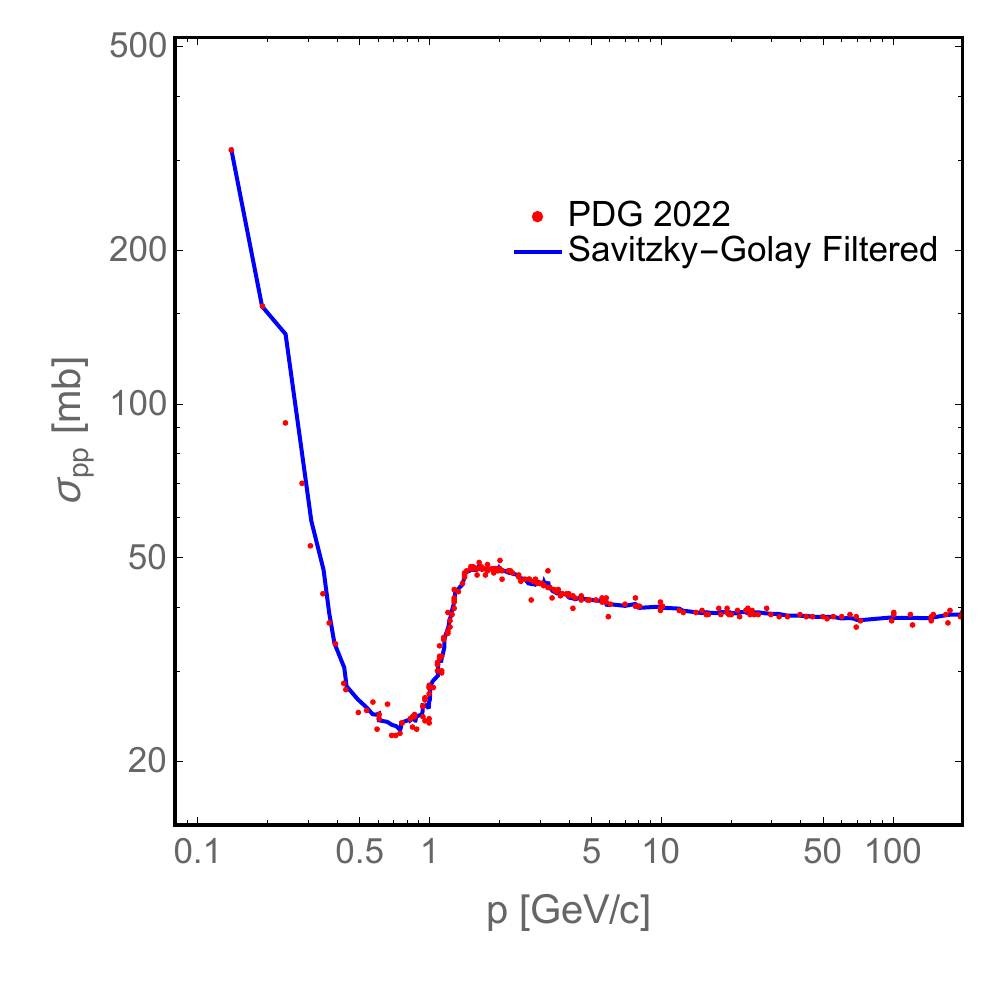}\qquad\includegraphics[width=6cm]{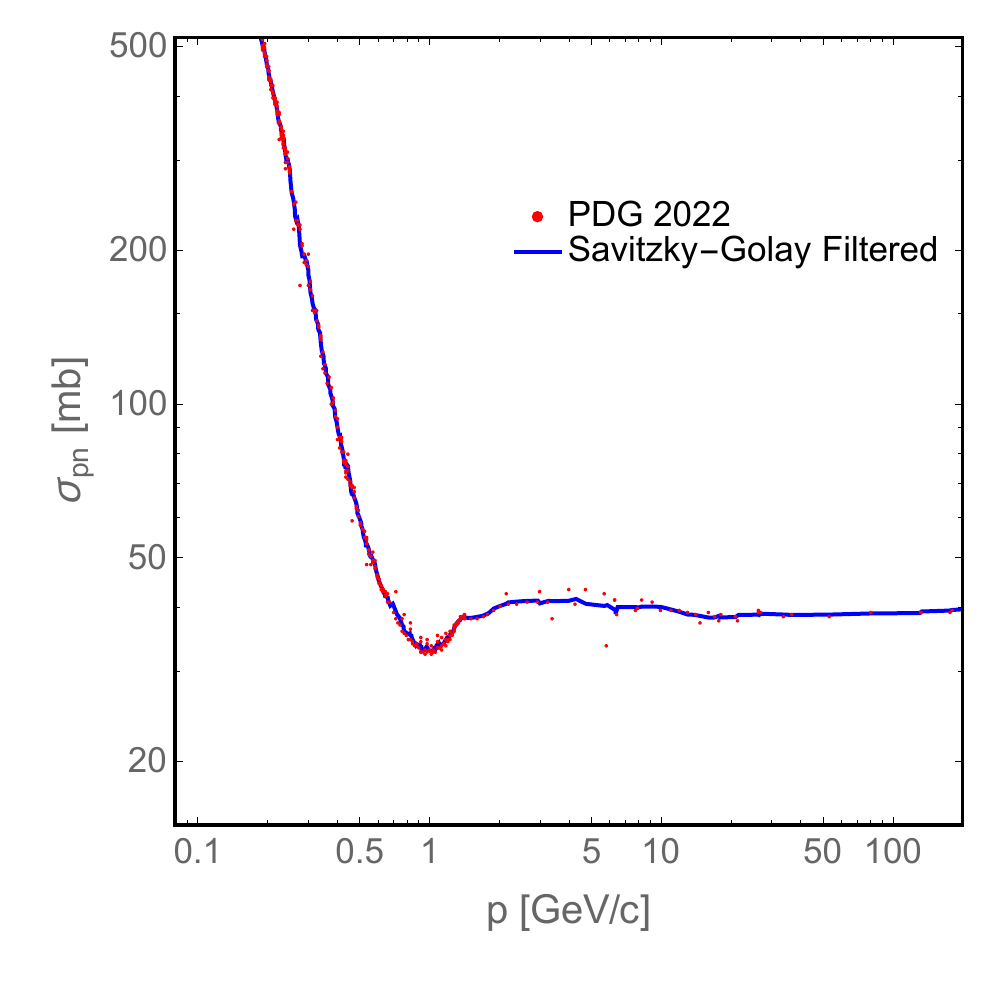}\qquad\includegraphics[width=6cm]{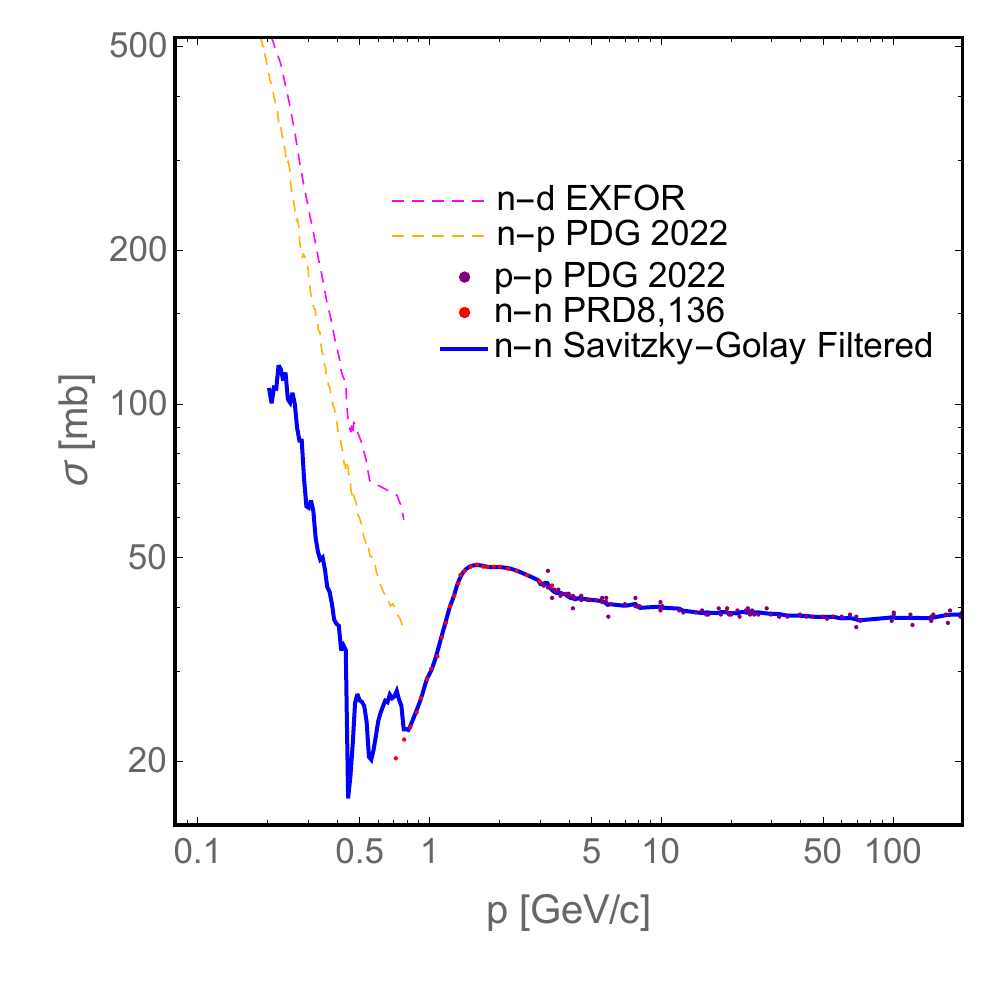}}

\vskip 2mm

\figcaption{7.5}{1}{Left: Proton-proton total cross section from the 2022 PDG compilation (red points; averaged when multiple measurements exist at the same beam momentum) together with the Savitzky-Golay smoothed curve (blue) adopted in this work. Middle: Same for the proton-neutron total cross section. Right: Neutron-neutron total cross section (blue curve) obtained from a Savitzky-Golay fit to three data sets: low-momentum points (orange dashed) extracted as $\sigma_\text{nd}-\sigma_\text{np}$, high-momentum points (purple) assumed equal to $\sigma_\text{pp}$, and intermediate-momentum data (red) taken directly from Ref.\cite{18}.}

\medskip

The nucleon-nucleon cross sections used in this work are shown in Fig.~1. Proton-proton and proton-neutron data are taken directly from the 2022 PDG compilation~\cite{16}; when several measurements exist at the same beam momentum we average them, and we apply a Savitzky-Golay filter~\cite{17} (window length 5, polynomial order 1) to reduce residual scatter. The lowest momentum measured is $\sim0.2$~GeV. We do not extrapolate below this value because, as shown later, hard scattering is already negligible compared with Bethe-Bloch energy loss in that regime.

Free neutron-neutron scattering cannot be measured directly. Following standard practice~\cite{18,19},$\sigma_\text{nn}$ is extracted by subtracting the neutron-proton cross section from neutron-deuteron data. For momenta $\lesssim0.8$~GeV multiple scattering inside the weakly-bound deuteron is small and the simple subtraction is accurate~\cite{20}. At high momenta ($\gtrsim3$~GeV) isospin symmetry implies $\sigma_\text{nn}\approx\sigma_\text{pp}$ up to tiny electromagnetic corrections, so we adopt the proton-proton values there. The intermediate region ($0.8-3$~GeV) is interpolated from the dedicated analysis of Ref.\cite{18}, with corrections from Refs.\cite{21,22,23}.
This piece-wise $\sigma_\text{nn}$ is displayed in the right-hand panel of Fig.~1.

Eventually the expected scattering number along the path is
\begin{equation}
\langle N\rangle=\sum_\textrm{T}\int dx\sigma_\textrm{hA/AA}\frac{\rho_\textrm{T}(h)}{A_\textrm{T}u}.
\label{Nexp}
\end{equation}

\medskip

We adopt the US Standard Atmosphere 1976~\cite{24} for the atmospheric density profile. Being independent of geographic location and season, it yields universal results; unlike more recent MSISE00 model~\cite{25}. The carbon dioxide level has risen from $0.0314\%$ in 1976 to about $0.0425\%$ in 2025~\cite{26}; we have updated the atmospheric composition accordingly. Because the microscopic interactions treated below are atomic, molecular species are first decomposed into their constituent atoms.

\vskip 4mm

\fl{2}\centerline{\includegraphics[width=6cm]{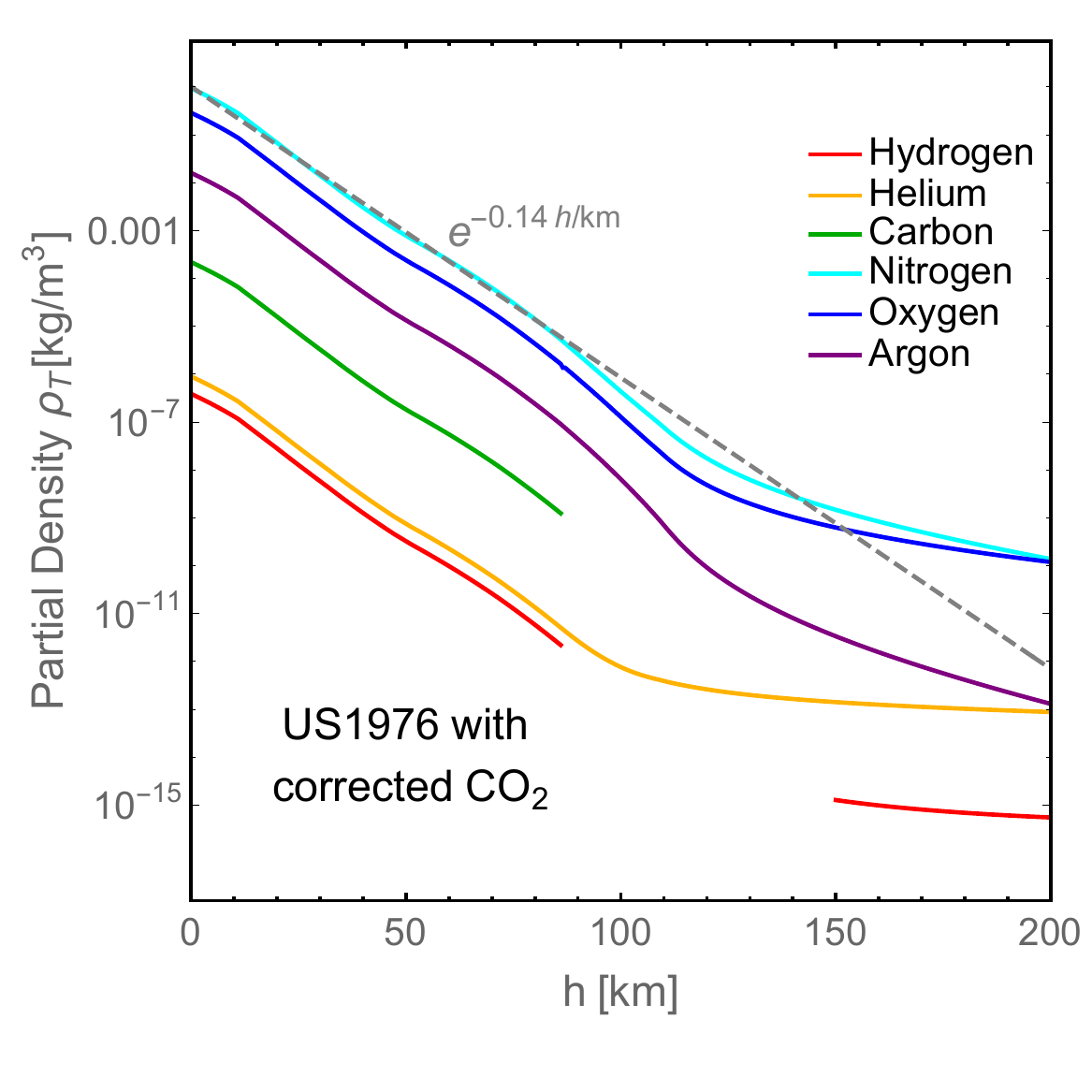}%\qquad\includegraphics[width=6cm]{spectraInd}
}

\vskip 2mm

\figcaption{7.5}{2}{Partial densities $\rho_\textrm{T}$ of the six dominant species-hydrogen (red), helium (orange), carbon (green), nitrogen (cyan), oxygen (blue) and argon (violet)-as functions of altitude in the updated US Standard Atmosphere 1976 model. The simple exponential fit $\rho_\textrm{T}\propto e^{-0.14h/\textrm{km}}$ (grey dashed) is shown for comparison. %Right panel: Improved factor of the exponential function fitting with altitude.
}

\medskip

Figure 2 shows the partial densities of the updated US Standard Atmosphere 1976. Below $86$~km the atmosphere is taken to be homogeneously mixed, so the relative abundances of nitrogen, oxygen, argon, carbon, helium and hydrogen are constant. Throughout this region each component follows an exponential profile, $\rho_\textrm{T}\propto e^{-0.14h/\textrm{km}}$, to excellent accuracy.

\vskip 4mm

\fl{3}\centerline{\includegraphics[width=6cm]{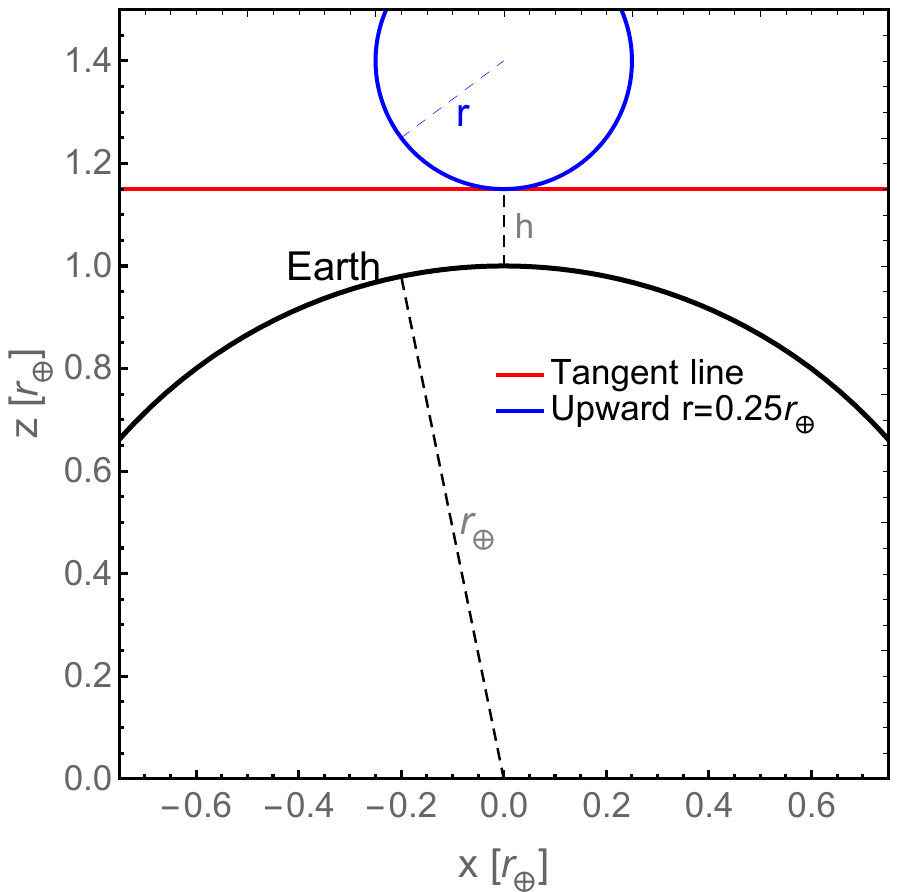}}

\vskip 2mm

\figcaption{7.5}{3}{The two trajectory geometries considered: a straight line tangent to the atmosphere (red) and an upward-bent trajectory with radius of curvature $r$ (blue).}

\medskip

{\it Results. Effect of perigee altitude.} The exponential fall-off of atmospheric density guarantees that both $\Delta\mathfrak{R}/\mathfrak{R}$ and $\langle N\rangle$ receive their dominant contributions from the segment closest to Earth. The natural control parameter is therefore the perigee altitude, $h$.

As a first baseline we consider a straight-line trajectory that is tangent to the atmosphere at altitude $h$ (impact parameter $b=r_{\oplus}+h$, with $r_{\oplus}=6371.2$~km); this is the red curve in Fig.~3. Although this path ignores geomagnetic bending, it captures the correct order of magnitude of the two stopping mechanisms.

\vskip 4mm

\fl{4}\centerline{\includegraphics[width=6cm]{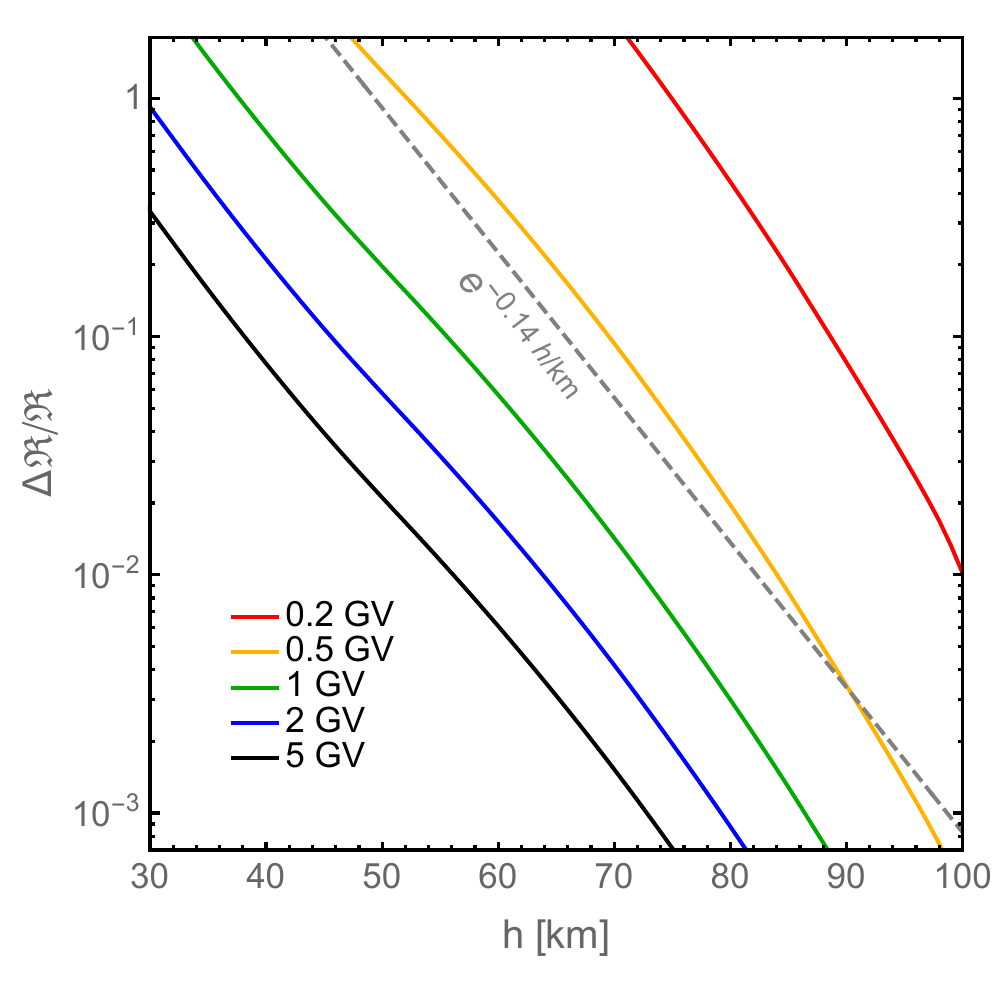}\qquad\includegraphics[width=6cm]{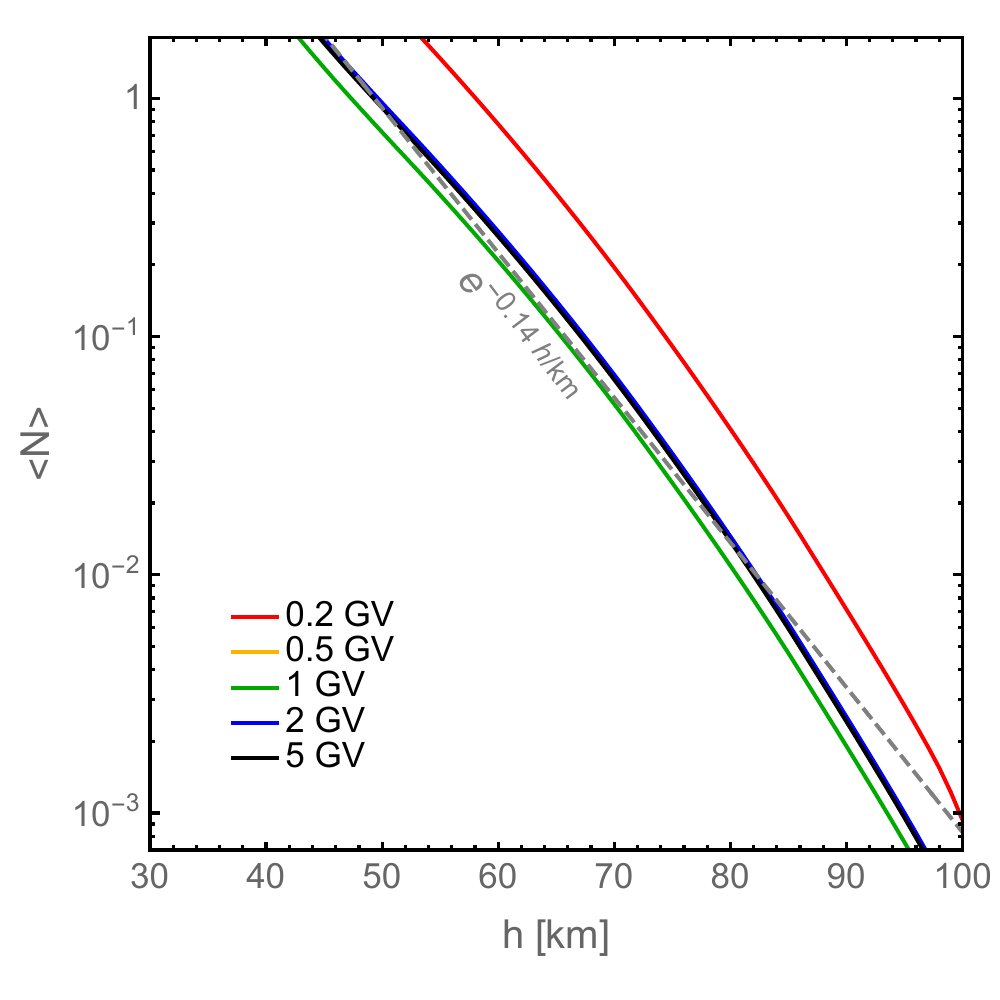}\qquad\includegraphics[width=6cm]{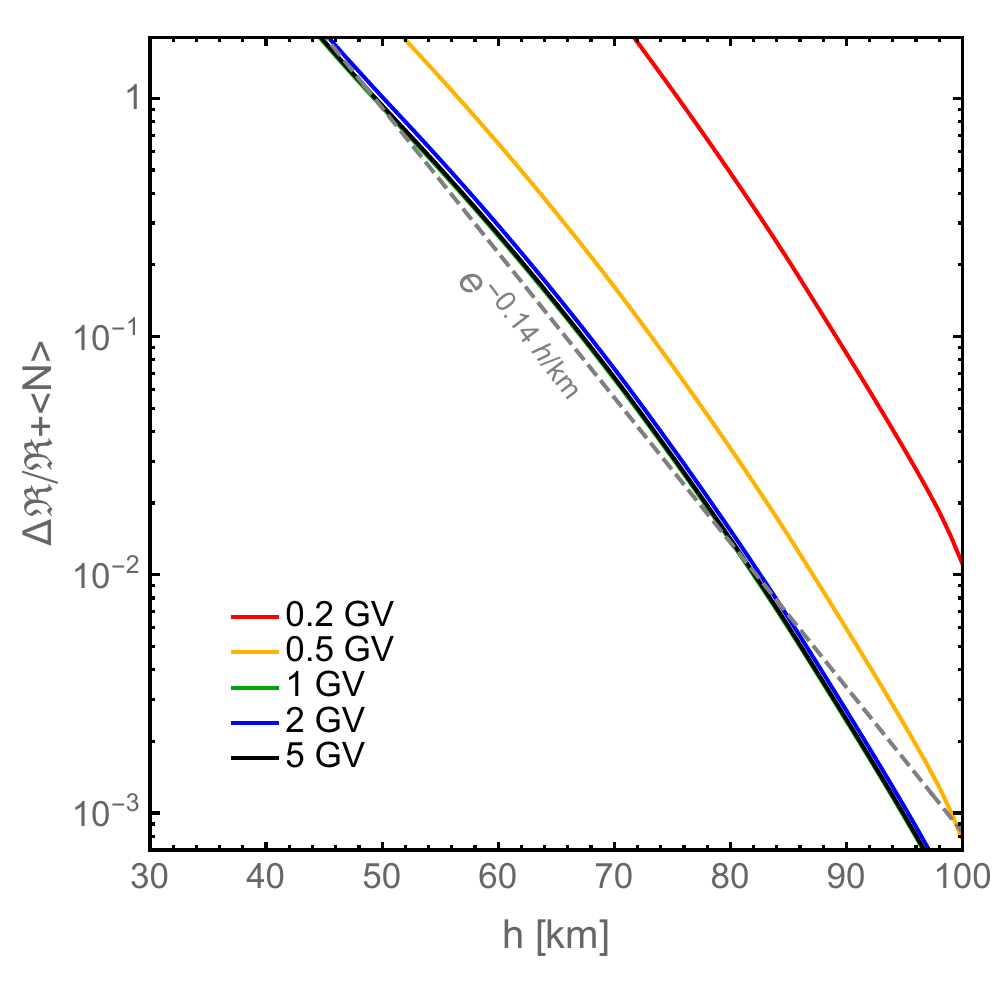}}

\vskip 2mm

\figcaption{7.5}{4}{The cumulative fractional rigidity loss $\Delta\mathfrak{R}/\mathfrak{R}$ (left), expected hard-scattering count $\langle N\rangle$ (middle) and their sum (right) versus perigee altitude $h$ for a straight-line tangent trajectory. Proton rigidities: 0.2 GV (red), 0.5 GV (orange), 1 GV (green), 2 GV (blue) and 5 GV (black). The altitude dependence is well described by the common exponential factor $\exp(-0.14\,h/\text{km})$ (grey dashed) and can therefore be factorized out.
}

\medskip

Figure 4 shows the integrated fractional rigidity loss $\Delta\mathfrak{R}/\mathfrak{R}$ (left), the expected hard-scattering count $\langle N\rangle$ (middle) and their sum (right) along an infinite straight-line trajectory tangent to the atmosphere for protons of five rigidities. Although $\Delta\mathfrak{R}/\mathfrak{R}$ and $\langle N\rangle$ describe distinct physical mechanisms, both are dimensionless measures of the breakdown of energy-conserving backtracing. We heuristically sum them to define a unified criterion for trajectory termination. All quantities exhibit the same $\propto e^{-0.14h/\textrm{km}}$ dependence on perigee altitude $h$ and can therefore be factorized out. 
The altitude range of $30-100$~km is chosen to highlight the variation around the critical perigee altitude $h$ where interactions become significant, which is much lower than the observer's altitude (e.g., $\sim400$ km for AMS-02).

{\it Effect of perigee curvature radius.} The straight-line tangent ignores bending by the Lorentz force; realistic trajectories must be obtained by (back-)tracing. Near perigee, however, any path is characterised by its local radius of curvature $r=\mathfrak{R}/B$ , almost independent of the remainder of the orbit. We therefore replace the tangent line by an upward-bending circular arc of radius $r$ that reaches the same perigee altitude $h$ (blue curve in Fig.~3). Downward-bending geometries with $r>r_{\oplus}+h$ can also yield perigee $h$, but are not considered here. The limit $r\to\infty$ recovers the straight-line case. The parameterized integrated trajectory is indeed a circle with radius $r$, while we have checked that the arc around the perigee almost contributes the whole integration. For simplicity the arc is taken to lie in the plane containing the Earth's centre and the perigee point.

\vskip 4mm

\fl{5}\centerline{\includegraphics[width=6cm]{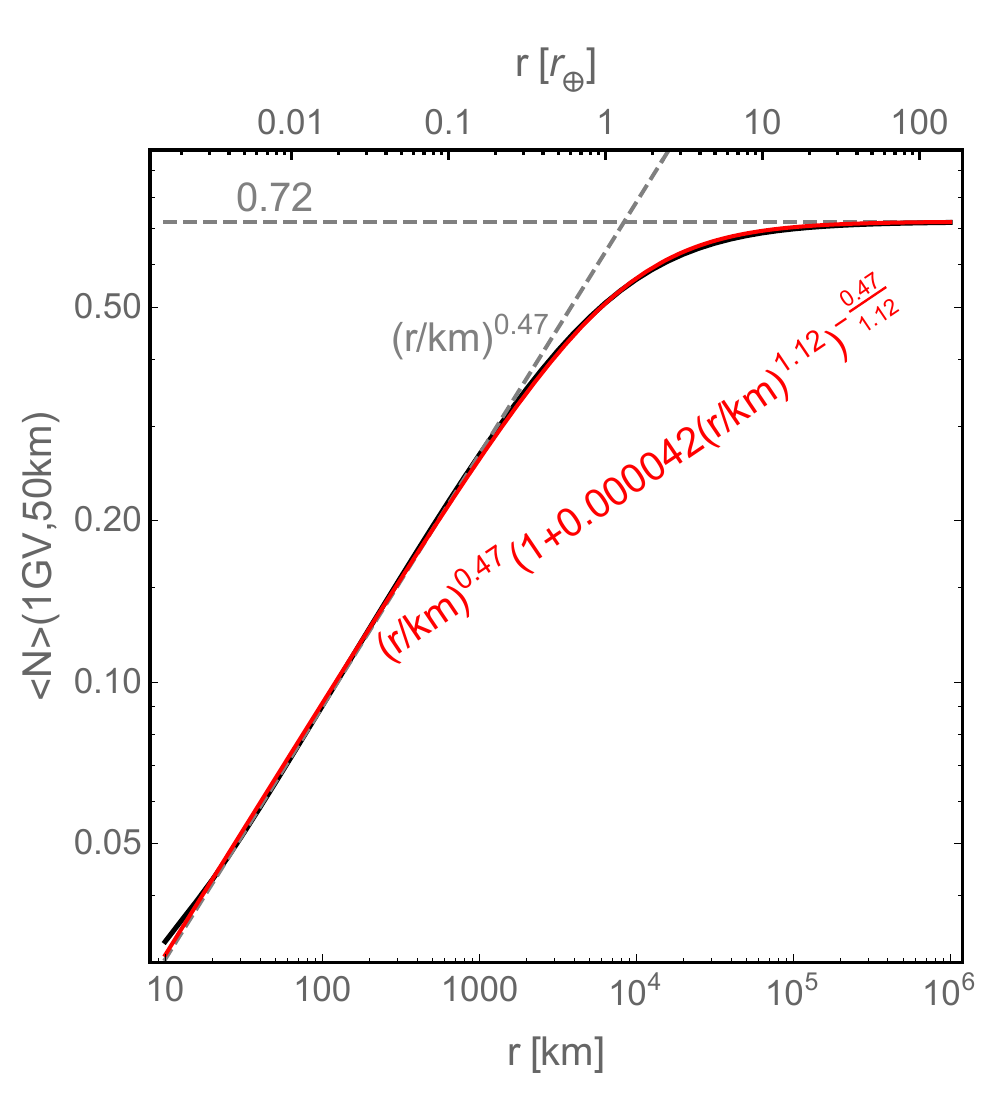}}

\vskip 2mm

\figcaption{7.5}{5}{The expected hard-scattering count $\langle N \rangle$ versus local radius of curvature $r$ for an upward-bent trajectory with perigee altitude $50$~km. The black curve gives the full calculation; grey dashed curves show the low-$r$ fit and high-$r$ asymptote; the red curve is the global fit.
}

\medskip

Figure 5 compares the numerical integration of $\langle N \rangle$ with its asymptotic and fitted forms. The behaviour of $\Delta\mathfrak{R}/\mathfrak{R}$ is identical. The global fit
\begin{equation}
g(r)=\Big(\frac{r}{\textrm{km}}\Big)^{0.47}~\bigg(1+0.000042\Big(\frac{r}{\textrm{km}}\Big)^{1.12}\bigg)^{-\frac{0.47}{1.12}},
\label{gr}
\end{equation}
normalised to its $r\to\infty$ limit, reproduces the calculation quite well. Combining this curvature factor with the altitude dependence gives the compact factorized forms, which works quite well if normalized with its $r\to\infty$ asymptotic value. With this fitting for local curvature radius $r$ around perigee and the above fitting for perigee altitude $h$, the two dimensionless variables can be factorized as
\begin{eqnarray}
\Delta\mathfrak{R}/\mathfrak{R}(h,r,\mathfrak{R},Z)&=&\Delta\mathfrak{R}/\mathfrak{R}(50\textrm{km},\infty,\mathfrak{R},Z)~e^{-0.14(\frac{h}{\textrm{km}}-50)}
~\frac{g(r)}{g(r\to\infty)},\\
\langle N \rangle(h,r,\mathfrak{R},Z)&=&\langle N \rangle(50\textrm{km},\infty,\mathfrak{R},Z)~e^{-0.14(\frac{h}{\textrm{km}}-50)}~\frac{g(r)}{g(r\to\infty)}.
\label{factorize}
\end{eqnarray}

\vskip 4mm

\fl{6}\centerline{\includegraphics[width=10cm]{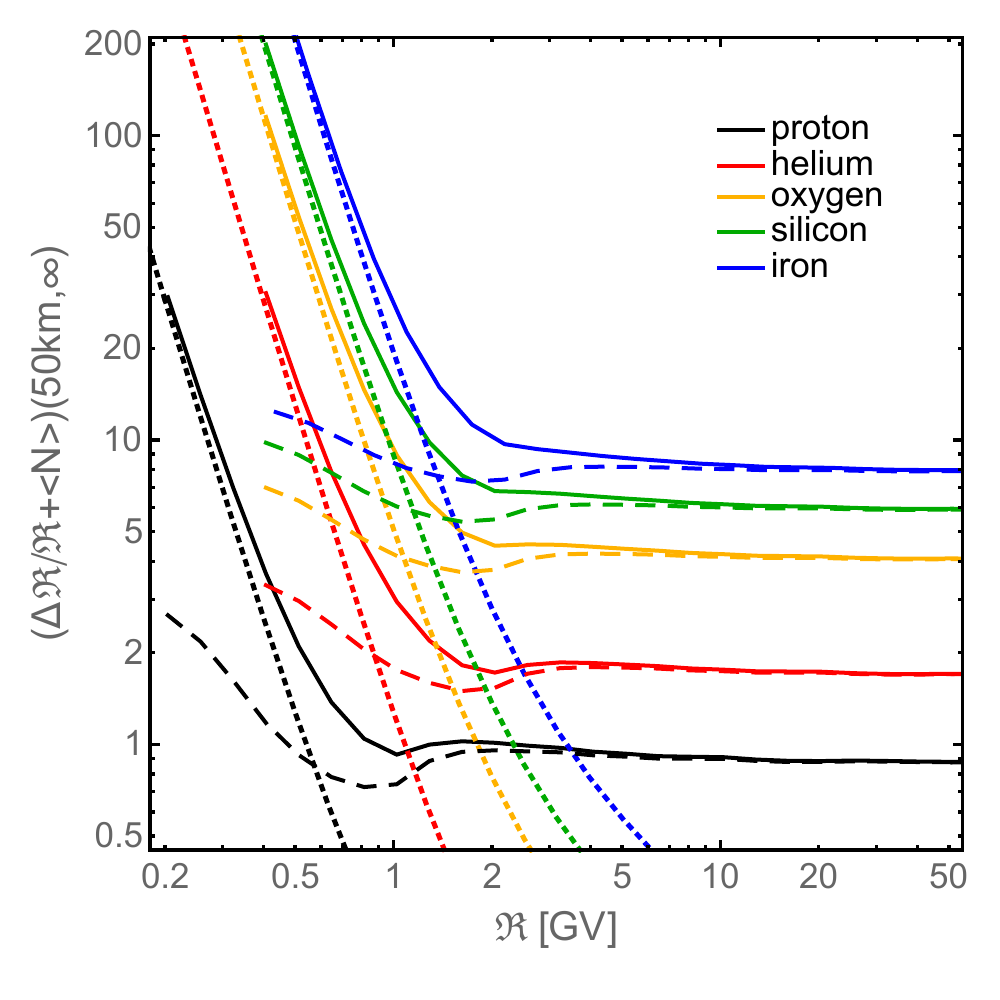}}

\vskip 2mm

\figcaption{7.5}{6}{The cumulative fractional rigidity loss $\Delta\mathfrak{R}/\mathfrak{R}$ (dotted), expected hard-scattering count $\langle N \rangle$ (dashed) and their sum (solid) versus incident rigidity for five elements-proton (black), helium (red), oxygen (orange), silicon (green) and iron (blue)-in a straight-line trajectory with perigee altitude $50$~km.}

\medskip

{\it Particle type and rigidity dependence.} Figure 6 shows the sum $\Delta\mathfrak{R}/\mathfrak{R}+\langle N\rangle$ versus rigidity for five cosmic-ray species on a straight-line trajectory tangent to the atmosphere at $h=50$~km. Bethe-Bloch energy loss, scaling approximately as $\beta^{-2}$, dominates at low rigidities, while hard scattering, with its nearly energy-independent cross section, takes over at high rigidities. The crossover occurs at $\sim0.57$~GV for protons and rises to $\sim1.4$~GV for iron. These reference values are inserted into Eq.~\ref{factorize} to obtain the full altitude- and curvature-corrected estimates. Again, values $\Delta\mathfrak{R}/\mathfrak{R}\gtrsim1$ obtained from the perturbative calculation are simply rescaled by Eq.~\ref{factorize} to the region $\Delta\mathfrak{R}/\mathfrak{R}\ll1$.

For protons at $50$~km, $\Delta\mathfrak{R}/\mathfrak{R}+\langle N\rangle\gtrsim1$ at all rigidities in the straight-line tangent geometry, so this altitude may serve as a reasonable lower boundary when a sharp boundary is imposed in backtracing. The chosen threshold of $\Delta\mathfrak{R}/\mathfrak{R}+\langle N\rangle$ should be smaller yet close to unity. Note, however, that the values in Fig.~6 neglect the finite-radius-of-curvature correction. For iron the sum is roughly ten times larger, so the same criterion requires the altitude-curvature factor to be $\sim0.1$, equivalent to raising the tangent altitude by $\sim16$~km.

{\it Discussion.} The physics-based termination criterion improves ad hoc backtracing boundaries. For a $1$~GV proton in the straight-line tangent geometry we obtain $\Delta\mathfrak{R}/\mathfrak{R}+\langle N\rangle=0.93$ at $50$~km, but $3.4$ at $40$~km. Primary charged cosmic rays therefore cannot originate from a perigee just above the formerly adopted $40$~km cutoff~\cite{3}. 

As a second illustration we again consider vertical incidence at $0^\circ$ longitude, $0^\circ$ latitude and 400 km altitude. Using the conventional algorithm of Ref.~\cite{2} (IGRF-13 field for epoch 2023 with the traditional $100$~km sharp boundary) the geomagnetic cutoff rigidity is $12.04$~GV. A very dedicated search shows that for $\mathfrak{R}=12.03809986288667$~GV the backtraced trajectory reaches a perigee of $98.3$~km before escaping to infinity. The old sharp-boundary prescription would discard this event, whereas the physics-based criterion developed here accepts it.

Because the physics-based criterion $\Delta\mathfrak{R}/\mathfrak{R}+\langle N\rangle$ is intrinsically probabilistic, the very notion of a penumbra becomes questionable. Penumbra denotes the narrow rigidity interval - usually around the vertical cutoff - where numerical backtracing switches repeatedly between ``allowed'' and ``forbidden'' as the rigidity is scanned in fine steps. Bethe-Bloch energy loss is deterministic, but hard scattering supplies only an \emph{expected} number of collisions; even when the path integral gives $\langle N\rangle=1$ there remains a finite survival probability. Thus the boundary between allowed and forbidden rigidities is unavoidably smeared, irrespective of whether the backtracing boundary is set at the physics-based $50$~km, or any other fixed altitude.

The allowed cone (i.e., the set of permitted incident directions for cosmic rays in LEO) can be approximated at high rigidities by the solid angle above the line of sight tangent to the sharp boundary~\cite{2}. For an arrival altitude of $400$~km, the use of a $20$~km boundary yields an allowed cone that is $1.0\%$ larger for protons ($1.5\%$ for iron) compared to a $50$~km ($66$~km for iron) boundary. Conversely, a $100$~km boundary yields a cone that is $1.7\%$ smaller for protons ($1.2\%$ for iron). Consequently, space dosimetry~\cite{6} is overestimated by approximately $1.0\%-1.5\%$. Furthermore, the AMS-02 experiment can utilize this additional $1.2\%-1.7\%$ of events vaively, which corresponds to billions of events given its 15 years of data collection.

A practical limitation of the present study is that the local radius of curvature around perigee is known only after a full backtracing calculation. In such a code the integrals for $\Delta\mathfrak{R}/\mathfrak{R}$ and $\langle N\rangle$ can be evaluated event-by-event, without recourse to the parametric factor $g(r)$ in Eq.~\ref{gr}, yielding the exact energy-loss and scattering probability for each trajectory and permitting any desired numerical threshold to be imposed a posteriori. This refinement in backtracing removes the uncertainties introduced by the sharp-boundary altitude approximation, enabling a single backtracing calculation to suffice for all cases.

\textit{Acknowledgements.} This work was supported by the National Natural Science Foundation of China (Grant No.~12505102,~12575119,~U2106201) and the Shandong Province Natural Science Foundation (Grant No.~ZR2024QA104).

\end{document}